\documentclass[pra,twocolumn]{revtex4}
\usepackage{epsfig}
\usepackage{graphicx}
\usepackage{amsmath}
\usepackage{natbib}
\usepackage{color}

\newcommand{\n}{\nonumber}
\newcommand{\bn}{\begin{eqnarray}}
\newcommand{\en}{\end{eqnarray}}
\newcommand{\eml}{\end{multline}}
\newcommand{\bml}{\begin{multline}}
\newcommand{\h}{\hspace}

\begin{document}

\title {Dynamical Resonances and Stepped Current in an Attractive Quantum Pump}
 \author{Kunal K. Das $^{1,2}$, Joshua Garner$^{1}$, and Kevin Ruppert$^{1}$}
\affiliation{$^{1}$ Department of Physical Sciences, Kutztown University of Pennsylvania, Kutztown, Pennsylvania 19530, USA}
  \affiliation{$^{2}$Department of Physics and Astronomy, State University of New York, Stony Brook, New York 11794-3800, USA}

\date{\today }
\begin{abstract}
We report on the transport properties of a single mode quantum pump that operates by the simultaneous translation and oscillation of a potential well.   We examine the dynamics comparatively using quantum, classical and semiclassical simulations. The use of an attractive or well potential is found to present several striking features absent if a barrier potential is used instead, as usually favored. The trapping of particles by the well for variable durations and subsequent release leads to a fractal-like structure in the distribution of the classical scattering trajectories.  Interference among them leads to a rich dynamical structure in the quantum current, conspicuously missing in the classical current. Specifically, we observe sharp steps, spikes and dips in the current as a function of the incident energy of the carriers, and determine that a dynamical version of Fano resonance has a role that depends on the direction of incidence and on multiple scattering by the potential.
\end{abstract}
%

\maketitle

\section{Introduction}

When precise and limited flow is a priority, time-varying pumps are a better alternative to static bias. Quantum pumps apply that principle to generate directed biasless flow at the scale of individual electrons and other carriers, but with quantum mechanics adding novel features absent in classical flow. Practically all quantum pumps studied have utilized barrier potentials, being a natural extension of classical analogs. But, as evident even in elementary treatments, quantum mechanics blurs the distinction between scattering by wells and by barriers, so that wells could be used just as well as barriers to generate flow. Wells, however, provide richer dynamics because of the inherent possibility of entrapment of carriers for varying durations, providing more diverse possibilities for mixing very different classical trajectories.  We identify and study some of those features in this paper in the context of a pump mechanism that mimics a traditional paddlewheel.

Quantum pumps were conceived by D. J. Thouless as a way of creating quantized flow based on the same topological arguments he used to explain the integer quantum Hall effect \cite{TKNN}, but by replacing a coordinate degree of freedom with time \cite{thouless, Niu-Thouless}. In the decades since, with advances in nanotechnology, the focus shifted to generating controlled directed flow by using time-varying potentials of charge \cite{brouwer-1,buttiker-floquet,turnstile,das-opatrny,avron-geometry-PRB,nanostructures,Kouwenhoven-PRL-1991,
entin-wohlman-SAW,Aono, arrachea-green, zhou-mckenzie,Makhlin-Mirlin-2001,Aleiner-Andreev-1998,Zhou-Spivak-Altshuler,kim-floquet}, spin \cite{chamon-spin,watson,blaauboer} and even entanglement \cite{samuelsson-buttiker,das-PRL} in typical mesoscopic circuits, and the scope broadened to include superconductors \cite{Giazotto-superconductor}, graphene \cite{Blaauboer-graphene} and carbon nanotubes \cite{Buitelaar-nanotube,buitelaar-nanotube-2}. However despite the sustained theoretical interest, experimental demonstration has been challenging, with varied success involving spin currents \cite{watson}, hybrid normal-superconducting systems \cite{Giazotto-superconductor} and carbon nanotubes \cite{buitelaar-nanotube-2}.

One of the major issues has been that the charge of the carriers, when subjected to time-varying potentials, creates competing effects that could potentially dominate \cite{switkes,brouwer-2}. This, along with progress in cold atom dynamics in microtraps and waveguides \cite{RMP-microtraps}, led one of us to suggest alternate implementation with trapped \emph{neutral} ultracold fermionic atoms  \cite{Das-Aubin-PRL2009}.  In a subsequent study \cite{Das-wavepacket}, we developed a novel approach for simulating general mesoscopic transport (including quantum pumps) with wavepackets of ultracold bosons, which can access details of transport dynamics at a single mode level instead of the usual multimode average inherent for electrons.

A series of studies \cite{Das-single-barrier,Das-double-barrier,Delos-Byrd,Das-PRE-pump,Das-paddlewheel-barrier} afterwards revealed that in this approach, the underlying scattering dynamics by a time-varying potential can be studied in its own right replete with rich features that include signatures of chaos, fractal structures, counterintuitive flow and offering, via semiclassical analysis, a fertile system for examining the interface between quantum and classical dynamics. This current study is aligned with these latter developments in the context of ultracold atoms, where the focus is on the details of the scattering dynamics involved rather than on the generation of net directed flow. But, differently from preceding studies, we examine for the first time the effects of using a well or attractive potential.

We describe the mechanism of an attractive paddlewheel pump in Sec.~I, and then present our wavepacket approach for examining its quantum dynamics in  Sec. II and classical dynamics in Sec. III. We discuss and analyze the primary features of the pumped current in Sec. IV, contrasting the classical and quantum results, and in Sec. V, we examine the scattered momentum distribution to explain those features. We use semiclassical simulations in Sec. VI to bridge the strikingly different results obtained in the quantum and classical scenarios. We conclude with a summary of our findings and possibilities for further work.

\begin{figure}[t]\includegraphics[width=\columnwidth]{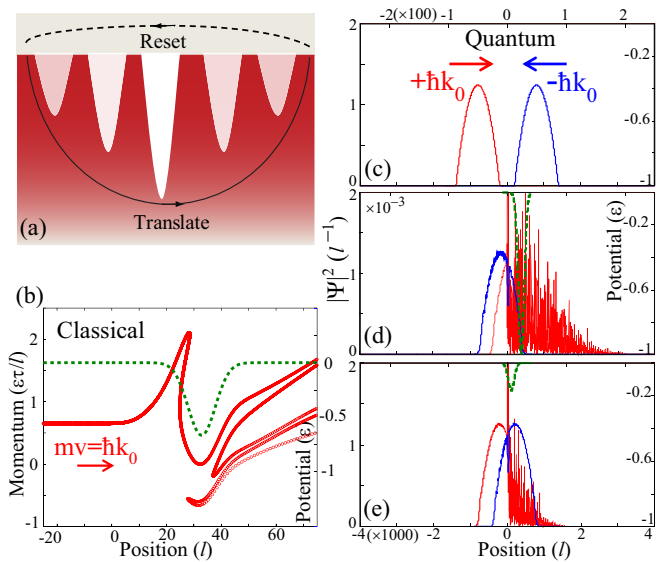}
\caption{(Color online) (a) The schematic of the attractive paddlewheel mechanism wherein a well potential oscillates while translating to the right and then resets. (b) A snapshot of a classical simulation with a stream of particles incident from the left, scattered by the potential. (c-e) Snapshots of a quantum simulation with counter-propagating wavepackets incident on the potential from either side and scattered by it. Note the packet incident from the right transmits almost unaffected.  In all cases the potential is shown as a dashed green line and gauged along the right axis, and for the quantum case (c-e), the length scale for the potential, along the upper axis, is expanded ($\times 10)$ for greater detail.
}\label{figure-1:schematic}
\end{figure}

\section{Attractive Paddlewheel pump}

A quantum pump in mesoscopic electronics generates directed flow without bias, by using time-varying potential through quasi one-dimensional (1D) nano-wires connected to macroscopic contacts that act as source and absorbing reservoirs for fermionic carriers.  Carrier motion \cite{Ferry-Goodnick} is assumed ballistic in the wires, so the current is determined by the scattering at the potential. The current can be defined as the integral over contributions at each incident momentum $p_0=\hbar k_0$,
\bn\label{fermion-current} J_F(t)&=&\int_{-\infty}^{\infty}\frac{dk_0}{2\pi}f(k_0) J(k_0,t)\\ J(k_0,t)&=&(\hbar/m) \langle
\psi(k,t)|k|\psi(k,t)\rangle/\langle \psi(k,t)|\psi(k,t)\rangle,\n\en
$f(k_0)$ is the Fermi distribution function and $\psi(k,t)$ the scattering wavefunction generated by incident mode $e^{ik_0x}$. The net current can be then understood as the sum over modes of the single mode current
\bn J_s(k_0,t)={\textstyle \frac{1}{2}}[J(+|k_0|,t)+J(-|k_0|,t)], \label{single-mode-current}\en the incoherent sum due to the lack of coherence between particles from different reservoirs and randomization of phase in each reservoir.  The single mode current will be of primary interest here, since it showcases the scattering dynamics prominently. Furthermore, we will assume implementation with ultracold atoms trapped in a quasi-1D waveguide, with focused lasers serving as the pumping potential \cite{Das-Aubin-PRL2009}.

Minimally, to operate as a continuous pump the potential needs to be cyclic, and it requires two independently varying parameters to generate sustained unidirectional flow. The simplest such pump mimics a paddlweheel, comprising of a single barrier that oscillates while translating a distance before resetting, just like the motion of a sequence of paddles in water. We explored such a pump for the case of a repulsive or barrier potential in a previous paper \cite{Das-paddlewheel-barrier}. The paddlewheel mechanism is therefore an obvious choice for an initial exploration of the effects of an attractive or well potential in quantum pumps. Assuming sinusoidal motion, an attractive paddlewheel pump can be implemented with the potential
\bn V(x,t)=U_0e^{-(x-f(t))^2/(2\sigma^2)}[1+A\sin(\omega_{osc} t +\phi)].\label{potential}\en
where $U_0<0$.  The depth of the well, oscillating with frequency $\omega_{osc}$, and its position, $f(t)=\mod(vt,d)$ resetting after distance $d=v\times2\pi/\omega_{tran}$, constitute the two independent time-varying parameters required. We set $\phi=3\pi/2$ and $\omega_{osc}=\eta\times\omega_{tran}$ with integer $\eta$, so for $\eta=1$ the bottom of the well traces a curve shown schematically in Fig.~\ref{figure-1:schematic}. As the well vanishes at $x=d$ at the end of a cycle, it re-emerges at $x=0$, the span $d$ chosen for the two frequencies to be commensurate.  Our choice of a Gaussian shape is dictated by its smoothness and for being the  typical laser profile, with a view to implementation with focused lasers in ultracold atoms.

In our simulations, the results are presented in dimensionless form, and in the case of ultracold atoms in a waveguide,  the transverse harmonic trap frequency $\omega_r$ can be used to set the energy, length and time units  $\epsilon=\hbar\omega_r$, $l=\sqrt{\hbar/m\omega_r}$ and $\tau=\omega_r^{-1}$, with $m$ being the mass of individual atoms. Such a choice of units yields a form of the Schr\"{o}dinger equation that is equivalent to setting $\hbar=m=1$. Unless otherwise specified, we use the parameters $U_0=-0.5,A=1,v=1.5,\sigma=5$ and $\omega_{osc}=0.2$, which sets the reset span of the pump to be $d=\eta\times 47.1$. These values were chosen to correspond to realistic experimental parameters \cite{Das-paddlewheel-barrier}.

\section{Quantum Simulation}

The single mode current can be simulated with wavepackets of ultracold atoms, confined to waveguides, with the group velocity simulating carrier motion \cite{Das-Aubin-PRL2009}.  Broad wavepackets, for which the time of passage through the potential substantially exceeds both the dwell time at the potential \cite{buttiker-landauer} and the time-periodicity of the potential, behave like plane waves as is usually assumed in mesoscopic transport. In order to emulate carriers incident from either side, wavepackets $\psi_\pm (x,t=0^-)$  initially centered to the left or the right of the  potential are propelled towards it with momentum $\pm \hbar k_0$ respectively. Each packet is allowed to evolve independently via the time-dependent Schr\"odinger equation for the time-varying pump potential until a time $T$ when the entire packet had interacted with the potential.  Figure~\ref{figure-1:schematic} shows snapshots of such a simulation.

The  scattered wavefunction $\psi_\pm(k,T)$ in momentum space, on integration yields the average current $J(\pm |k_0|,T)=\int dk\ k |\psi_\pm(k,T)|^2$ and scattering probabilities $\int_0^\infty dk |\psi_\pm(k,T)|^2$ and $\int_{-\infty}^0 dk |\psi_\pm(k,T)|^2$. In experiments with cold atoms, wavepackets can be prepared within an additional axial trap centered away from the potential, and transport initiated by switching it off and giving the atoms the appropriate momentum with Bragg beams \cite{Ketterle}, $\psi(x,0^+)=e^{\pm ik_0x}\psi(x,0^-)$, and after scattering, image the spatial and momentum distribution.

The asymmetric scattering by the time-varying potential results in non-vanishing net current $J_s(k_0,t)$ on averaging the contributions from the left and right incident packets according to Eq.~(\ref{single-mode-current}). The multimode current of fermions can be computed by sampling the single-mode current over the relevant range of incident momenta and approximating the integral in Eq.~(\ref{fermion-current}) by a Riemann sum. We will not focus on that here, since that would only average \cite{Das-paddlewheel-barrier} out many of the prominent features  of the scattering dynamics that is of primary interest here. In our simulations, we leave out contributions from outgoing carriers, corresponding to packets moving away from the potential at $t=0$, since their interaction, if any, with the potential is negligible and their contributions to the net current are therefore typically small \cite{Das-wavepacket}.

\section{Classical Simulation}

We also simulate the dynamics of the pump classically to distinguish features that are of strictly quantum origin. The wavepackets are replaced by a stream of uniformly-spaced particles matching the initial momentum of the quantum wavepackets $p_0=\pm \hbar k_0$.  The particles are propagated by Hamilton's equations,
\bn \frac{dx}{dt}=\frac{\partial H}{\partial p};\h{2mm}  \frac{dp}{dt}=-\frac{\partial H}{\partial x},\label{classical-equations}\en
until their interaction with the well is effectively complete.  The particles are then assembled in momentum bins of equal width $[p_i-\delta p/2, p_i+\delta p/2]$, and a histogram of the number of particles $N_i$ in each bin is the classical counterpart of the final momentum distribution. The total current is then calculated to be the average of the momentum distribution, weighted by the value of the momentum, arising from identical streams incident from the left ($p_0=+\hbar k_0, x_0<0$) and from the right ($p_0=-\hbar k_0, x_0>0$) of the potential,
\bn J_{\pm}=\frac{1}{N}\sum N^\pm_i p_i\h{1cm} J={\textstyle \frac{1}{2}}(J_++J_-).\en

\begin{figure}[t]\includegraphics[width=\columnwidth]{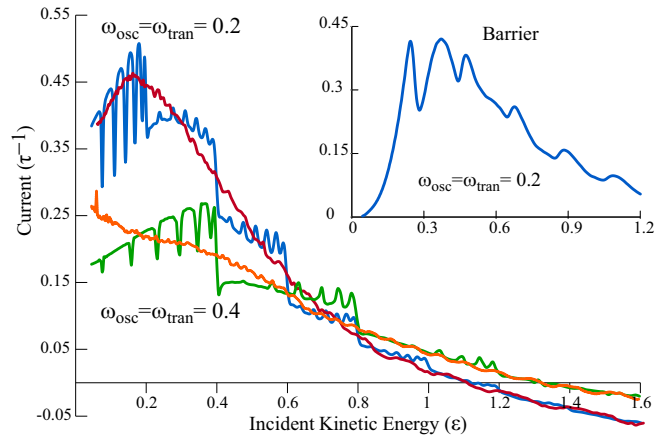}
\caption{(Color online) The pumped current as a function of the kinetic energy of the incident packet, for a well paddlewheel, is plotted for both classical and quantum dynamics. The quantum mechanical current shows three features (i) steps at multiples of the oscillation frequency, (ii) sharp dips over the first step and (iii) spikes over subsequent steps. These features are conspicuously absent in the classical current which however follows the general trend of the quantum current. The current profile is displayed for two different periodicities $\omega_{osc}=\omega_{tran}=0.2$ and $0.4$, the latter showing the proportionate doubling of the step periodicity. The inset shows the quantum current for a barrier paddlewheel, with otherwise identical parameters.
}\label{figure-2:main}
\end{figure}

\section{Features of Pumped Current}

We will examine the features of the pumped current, both classically and quantum mechanically, plotting both together for each case and aspect we consider. We find that the quantum current and scattering profile display prominent features conspicuously missing in their classical counterparts, which however always follow the general trend of the former as a sort of average or coarse-graining of the quantum picture. The quantum pumped current for a paddlewheel with a well-potential is plotted in Fig.~\ref{figure-2:main} as a function of incident kinetic energy for two cases, both with $\omega_{osc}=\omega_{tran}$. Three features are immediately prominent: (1) There are sharp stepped reductions of the current when the incident kinetic energy is a positive integer multiple of the oscillation period, $k_0^2/2=n\times \omega_{osc}$. (2) For the first step, when $ k_0^2/2<\omega_{osc}$, sharp dips appear in a pattern that bring the dips progressively closer together. (3) At subsequent steps $k_0^2/2 >\omega_{osc}$, the dips are replaced by spikes in a similar recurring pattern. Notably, these features are absent in the current generated by the equivalent barrier paddlewheel as is shown in the inset in Fig.~\ref{figure-2:main}, for which $U_0\rightarrow -U_0=+0.5$, all else remaining the same \cite{Das-paddlewheel-barrier}.  The sharp dips and spikes are entirely absent, and instead of the step structure there is a smooth undulation at the periodicity of the oscillation.  This establishes that all of these features arise specifically from having an attractive pumping potential.

\begin{figure}\includegraphics[width=\columnwidth]{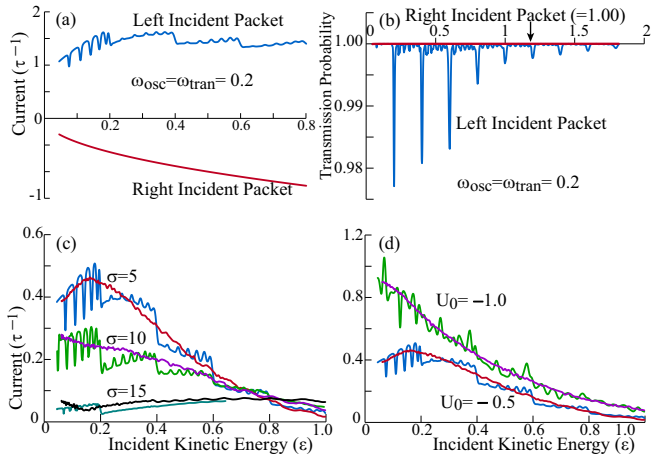}
\caption{(Color online) (a) The distinct features of the quantum current originate entirely in the scattering of the packet incident from the left of the potential, further confirmed by plotting the transmission probability for packets incident from both directions in (b) which shows that the packet incident from the right transmits completely. The parameters for panels (a) and (b) correspond to the $\omega_{osc}=\omega_{tran}=0.2$ quantum current profile in Fig.~\ref{figure-2:main}.
(c) Increasing the well width, $\sigma=5$ to $10$ and $15$ reduces the current, but retains its features, whereas (d) increasing the well depth, $U_0=-0.5$ to $-1.0$ increases the current but smudges its features. For all cases, the classical current follows the average trend of the quantum current, but lacks its distinctive features.
}\vspace{-2mm}\label{figure-3:Width_Depth}
\end{figure}

The features arise only from the left incident packet as shown in Fig.~\ref{figure-3:Width_Depth}(a) where we plot separately the contributions to the current due to wavepackets incident from the left $J(+|k_0|)$ and from the right $J(-|k_0|)$. This is further underscored in Fig.~\ref{figure-3:Width_Depth}(b) where the transmission probability due to the respective packets are plotted, and there is 100$\%$ transmission for incidence from the right, whereas there are sharp dips in the transmission probability for incidence from the left. The latter corresponds to the primary features in the current profile with the largest dips in transmission occurring where the steps appear in the current.  Although present in the probability distribution, those features are magnified by the momentum redistribution that also contributes to the current.

We now show that the dips arise from a mechanism akin to Fano resonances \cite{Fano,RMP-Fano}, which occur when the incident energy matches a bound state energy $E_b<0$, but now subject to Floquet's theorem whereby the scattered state will display sidebands $E_n=E_0\pm n\omega$, with $n=0,\pm 1,\pm 2, \cdots$ and $\omega=\omega_{osc}=\omega_{tran}$ in this case.  Such resonances have been noted in some previous studies for scattering by an oscillating barrier \cite{Haavig-Reifenberger, Reichl}, but for a paddlewheel there is simultaneous translation as well, which not only allows for net current to be generated in a biasless circuit, but also fundamentally alters the dynamics. Thus, while for simple oscillation $E_0=k_0^2/2$, in our case for wavepackets incident from the right and traveling to the left with negative group velocity ($k_0=-|k_0|$), we have instead $E_0=(|k_0|+v)^2/2$, accounting for the relative velocity of packet and potential. Therefore, resonance with a bound state will require
\bn   |k_0|=-v\pm \sqrt{2(-|E_b|+n\omega_{osc})}\label{res-cond-negative}\en
With our choice of parameters, where $v=1.5$ and $\omega_{osc}=0.2$ and for the \emph{median} depth of the well, corresponding to $U_0=-0.5, A=0$, the bound state eigenvalues are $E_b=$(-0.433,-0.307, -0.198, -0.108, -0.0405, -0.000247). It is therefore clear that the condition in Eq.~(\ref{res-cond-negative}) cannot be satisfied except for high values of $n$, with proportionately lower probability, else the right hand side would be negative or complex. Hence, the transmission probability for packets incident from the right is featureless.
In the case of the packets incident from the left, the incident velocity $k_0=+|k_0|$ is positive, and $E_0=(|k_0|-v)^2/2$, accounting for the relative velocity of packet and potential, and the resonance with a bound state now requires
\bn |k_0|=v\pm \sqrt{2(-|E_b|+n\omega_{osc})}\label{res-cond-positive}\en
This can however be satisfied for a range of values starting from the lowest value of $n=1$, and therefore the packets incident from the left show striking features associated with Fano-like resonance with bound states.

\begin{figure}[t]\includegraphics[width=\columnwidth]{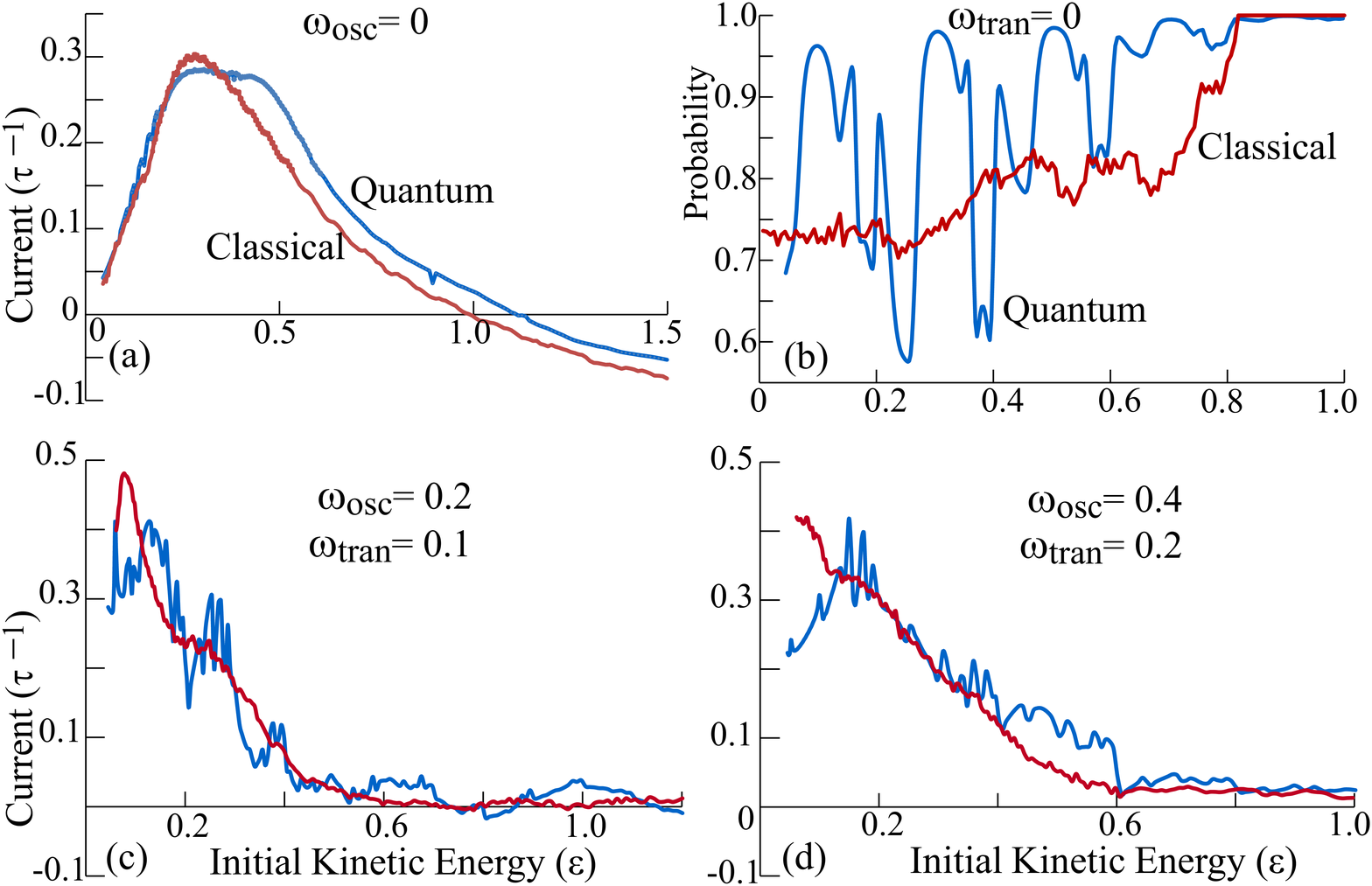}
\caption{(Color online) The distinct features in the quantum mechanical current are absent if \emph{either} of the two frequencies is set to zero, (a) $\omega_{osc}=0$ or (b) $\omega_{tran}$=0, indicating that they arise due to the interplay of the two periodicities. Those features are  diminished if the two frequencies differ, although still commensurate; two cases of $\eta=\omega_{osc}/\omega_{tran}=2$ are shown (c) $\omega_{osc}=0.2,\omega_{tran}=0.1$ and (d) $\omega_{osc}=0.4,\omega_{tran}=0.2$. For all cases, the classical current follows the average trend of the quantum current, but is relatively smooth.
}\vspace{-2mm}\label{figure-4:Frequency}
\end{figure}

The key difference between Eqs.~(\ref{res-cond-negative}) and (\ref{res-cond-positive}) is the sign of the velocity $v$, not its magnitude, and the sign is determined by the direction of incidence.  This is consistent with the dramatic difference between left and right incident packets, where despite the higher relative velocity, the right incident packets transmit without any features as seen in Fig.~\ref{figure-3:Width_Depth}(a,b). Furthermore, these equations also highlight the importance of the oscillation, since without it the square-root term would be imaginary in both cases, making a resonance condition impossible, and this is confirmed in Fig.~\ref{figure-4:Frequency}(a), where we switch off the oscillation and the resonance structure completely disappears.

The dynamics is however further complicated by repeated interaction of the wavepacket with the well potential as it resets in position after each cycle.  This is particularly relevant for lower incident velocities for which the time of traversal is longer. Repeated interactions can enable reaching higher sideband values $n$, and hence the prominent dips at low incident energy.

\begin{figure}[t]\includegraphics[width=\columnwidth]{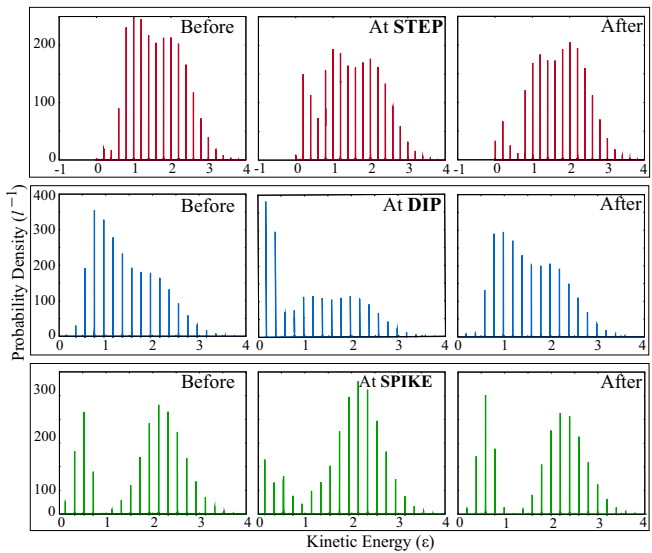}
\caption{(Color online) Scattered energy distribution for a wavepacket incident from the left, as a function of the kinetic energy $k^2/2$, shown left to right for `\emph{Before}' (less than),  `\emph{At}' and `\emph{After}' (greater than) one of the values of $k^2/2$ where each of the following features appears in the quantum mechanical current in Fig.~2  (for $\omega_{osc}=\omega_{tran}=0.2$):  Step (top panels), Dip (middle panels) and Spike (bottom panels)
}\vspace{-5mm}\label{figure-5:momentum}
\end{figure}

We examined the current profile by varying several of the relevant parameters. In Fig.~\ref{figure-3:Width_Depth}(c) we varied the well width and in Fig.~\ref{figure-3:Width_Depth}(d) we varied the well depth. It is clear that the features are left intact on increasing the well width, but the net pumped current diminishes. On the other hand, increasing the well depth increases the pumped current as should be expected from a stronger potential, however the features are diminished.

In Fig.~\ref{figure-4:Frequency}, we illustrate the role of the two frequencies at play, by turning off each in turn. When we set $\omega_{osc}=0$ in Fig.~\ref{figure-4:Frequency}(a), the well translates with fixed median depth $U_0=-0.5$, and then resets, and in this case, all the manifestly quantum features disappear, and the current profile closely follows the classical current. Then we set $\omega_{tran}=0$, in Fig.~\ref{figure-4:Frequency}(b) whereby the well oscillates in place without translating.  Due to the symmetry there is no net current \cite{das-opatrny}, so we plot the transmission probability due to the left incident packet and find substantial undulations, but quite distinct from when both periodicities are present. It is therefore clear that the quantum features arise as an interplay between the two frequencies. Furthermore, it is also essential to have the two frequencies be the same, not just commensurate, as we checked by doubling the ratio of the two frequencies to be $\eta=2$ in Figs.~\ref{figure-4:Frequency}(c) and (d). We show two cases $\omega_{osc}=0.2,\omega_{tran}=0.1$  and $\omega_{osc}=0.4,\omega_{tran}=0.2$, and for both, while there are oscillations, the features are not quite as prominent and are rather irregular. This further supports that the interplay of the two frequencies is crucial for creating the striking features seen in Fig.~\ref{figure-2:main}.

\section{Momentum Distribution}

In order to better understand the features of the quantum current, we examine the scattered momentum space probability density, plotted as a function of the outgoing kinetic energy in Fig.~\ref{figure-5:momentum} for a left incident packet. We do so for a representative case for the three principal features: step, dip and spike, and discuss them below:

\emph{Steps}: At the steps, a new Floquet peak appears in the scattered distribution around zero kinetic energy, indicating that as the incident energy reaches the oscillation energy $\omega_{osc}=\omega_{tran}$, this new lower value becomes energetically available and signals a shift of the distribution to lower energies.  This causes the sudden reduction in the net current manifest as a step structure. Notably, even for a barrier potential, undulations in the current profile is observed at a period corresponding to that of the oscillation. However, it is much more dramatic in the attractive case, since there is a dip in the transmission as well, as seen in Fig.~\ref{figure-3:Width_Depth}(a), signifying that a fraction of the particles become trapped in the well for a while when the incident kinetic energy matches the energy of oscillation.

\emph{Dips}: At the dips, we see a bimodal distribution with a noticeable enhancement of the peaks near zero energy. This suggests a resonant behavior where the energy of the incident packet matches a bound state and is trapped for a while during the cycles and eventually escapes with reduced energy.

\emph{Spikes}: The momentum distribution shows that in the regions between the spikes, there is a clear bimodal distribution, whereas at the spike itself the two parts blend with a skewing to higher momenta. This is just the reverse of what happens at the dips where the bimodal distribution becomes most pronounced, which suggests that spikes actually represent a narrower regime between broader, shallower resonance dips, and the same mechanism is at work as at the much sharper dips.

\section{Semiclassical Analysis}

\begin{figure}[t]\includegraphics[width=\columnwidth]{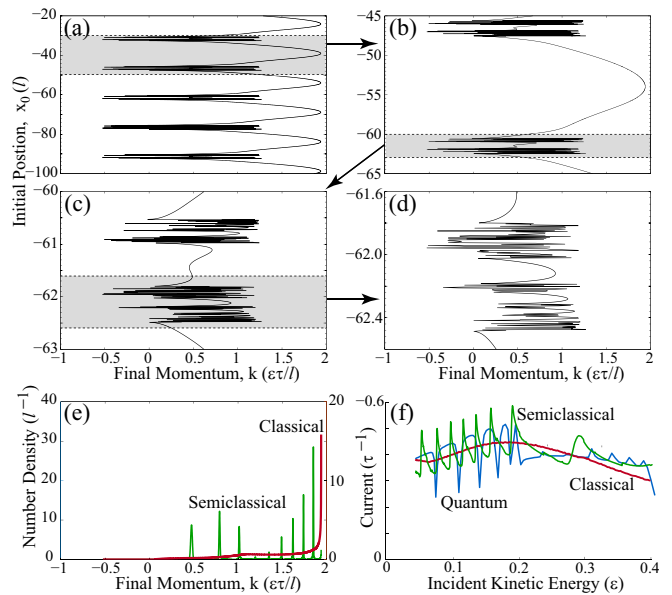}
\caption{(Color online) (a)-(d) Plot of the initial position of the classical line of particles versus their scattered momenta for a specific incident momentum of $k_0=+0.475$, displayed at various levels of zoomed-in detail, with (a) showing the periodicity, and the rest showing details within each period. (e) Semiclassical (green spikes, left axis) and classical (smooth red, right axis) scattered momentum distribution, both obtained from evaluating $\psi_{SC}$ in Eq.~(\ref{semi-classical}) by including and excluding interference among branches, respectively; the Floquet peaks emerge from the interference. (f) Comparison of the semiclassical with the classical and quantum current profiles for the case $\omega_{osc}=\omega_{tran}=0.2$ displayed in Fig.~\ref{figure-2:main}.
}\label{figure-6:semiclassical}
\end{figure}

We conducted a semiclassical analysis to get some insights into the contrast between the quantum and the classical currents.  In addition to Hamilton's equations in Eq.~(\ref{classical-equations}), we need to  time evolve $\tilde{S}(p,t)$ the momentum space counterpart of the classical action,
\bn \frac{d\tilde{S}}{dt}=-x\frac{d p}{d t}-H,\en
which serves as the phase in the semiclassical wavefunction in momentum space,
\begin{equation}
\psi_{SC}(p,t)={\textstyle \sum_j}\sqrt{\rho_0(x^j_0)}\left| \partial p/\partial x_0 \right|_{x^j_0} ^{-\frac{1}{2}}e^{i[\tilde{S}_j(p,t)/\hbar-\mu_j\pi/2]}\label{semi-classical}
\end{equation}
with $\mu_j$ being the Maslov index \cite{Maslov,Das-single-barrier}. Since the potential is periodic, the final momentum $p\equiv\hbar k$ is a periodic function of the initial position, $x_0$ of the particles, with each $p$ having contributions from multiple initial positions $x_0(p,t)$, as evident in Fig.~\ref{figure-6:semiclassical}(a), where a vertical slice corresponding to a given momentum would intersect the curve at multiple points.  It is clear from that figure, that even within each period there are numerous branches arising from particles incident at the well at different segments of a cycle that still lead to the same final momentum. Therefore, the sum in Eq.~(\ref{semi-classical}) is over both (i) intra-cycle branches within a cycle and (ii) inter-cycle repetition due to the periodicity. In the semiclassical wavefunction, the initial density of particles determines $\rho_0$ chosen to match the profile of the initial quantum probability distribution, the local slope in Fig.~\ref{figure-6:semiclassical} (a) of $x_0$  versus $p$ determines the Jacobian $\left| \partial p/\partial x_0 \right|$, whereas $p$ versus \emph{final} position $x$ (not plotted) fixes the Maslov index as follows: Within each cycle, we start from $\mu_j=0$, and at each turning point where $dp/dx=0$, it is incremented by $+1$ or $-1$ for clockwise or counterclockwise turns respectively \cite{Das-single-barrier}.

Our conclusions however can only be qualitative in nature, because as shown in Fig.~\ref{figure-6:semiclassical}(a)-(d), the attractive paddlewheel contains a fractal-like structure of branches in the scattered momenta, making it challenging to evaluate the semiclassical wavefunction precisely. We had to, of a necessity, approximate the Maslov index using the initial position versus $p$ plot, because the final position versus $p$ is substantially more complicated due to branches getting twisted and stretched with time. We also introduced cutoffs to limit the number of branches.

Figure~\ref{figure-6:semiclassical}(e) shows the \emph{semiclassical} momentum distribution resulting from retaining only the two major branches in each cycle, where $|\psi_{SC}(p,t)|^2$ contains the effect of interference among the different classical trajectories that creates the Floquet peaks seen in the quantum distribution. Also shown is the \emph{classical} momentum distribution obtained by taking the absolute-square of each term in Eq.~(\ref{semi-classical}) \emph{before} summing, whereby the phase information is lost and the Floquet peaks are replaced by a relatively smooth curve.

In Fig.~\ref{figure-6:semiclassical}(f), we plot the semiclassical current $\int dp\ p|\psi_{SC}(p,t)|^2$ for a range of incident momenta. We see emergent features at about the same locations as where quantum mechanical current profile has structure.  Since the only difference between the classical and the semiclassical cases is that the different trajectories with the same final momentum can interfere for the latter, it is clear that the quantum mechanical current acquires its prominent structures mostly due to such interference, caused by particle density trapped in the well for varying durations and subsequently released with a range of different momenta.
Considering that our cutoffs leads to inclusion of only the major branches in each cycle, it is to be expected that our computed semiclassical current is not a close quantitative match for the quantum mechanical current. However, progressive inclusion of more branches should improve the agreement, with increasingly finer branches having diminishing contributions.

\section{Conclusions}

We studied a quantum pump that operates with a single well or attractive potential executing a cyclical motion that mimics a traditional paddlewheel.  In contrast to pumps operating by barrier or repulsive potentials, the single mode pumped current demonstrates a set of striking features that include sharp dips, spikes and steps as a function of the incident carrier momentum.  We showed that the dips arise from a dynamical version of Fano resonance when the incident energy matches a bound state energy of the well, but with the important caveat that in our case, the relative velocity of the carriers and the well creates a significant asymmetry between incidence from  the left and the right. Multiple interactions with the potential during time of traversal also impact the features allowing access to higher order Floquet sidebands creating deep resonances. Examination of the momentum distribution and a qualitative semiclassical analysis confirmed that the trapping of particles by the well for variable durations lead to a very rich and a fractal-like structure of the scattered momenta.  Interference of all such trajectories having the same final momenta creates the prominent features in the quantum current, which are conspicuously absent in the classical current, also simulated here in conjunction.

From an applications standpoint, the sharp steps and dips visible in the current could be used to control the flow precisely by tuning the parameters to be in their vicinity.  There are substantial possibilities for future research on the quantitative analysis of the semiclassical dynamics to examine how the interference of the classical trajectories at different hierarchy of branch size inclusions approach the quantum dynamics.  The difference between classical and quantum dynamics is clearly magnified by the use of a well potential, and hence this can be a useful system to examine the interface and crossover between quantum and classical dynamics, as well as in the study of quantum chaos \cite{Delos-Byrd}. Well-based pumps clearly have much richer dynamics, and therefore, more complicated variations like a turnstile \cite{Das-double-barrier,turnstile} would be interesting to examine in future works, particularly in regards to if and how the single-mode dynamical features are manifest in the context of multimode fermionic systems.

\section{Acknowledgment}
We acknowledge discussions with S. Aubin, and the support of the NSF under Grants No. PHY-1313871 and PHY-1707878.

\vfill


\end{document}